# An Existence Criterion for Low-Dimensional Materials

Jiapeng Chen[1], Biao Wang[2, 3,*], Yangfan Hu[3]

[1]School of electronics and information technology, Sun Yat-Sen University, 510275 GZ, China
[2]School of physics, Sun Yat-Sen University, 510275 GZ, China
[3]Sino-French Institute of Nuclear Engineering and Technology, Sun Yat-Sen University, 510275 GZ, China
*Corresponding. wangbiao@mail.sysu.edu.cn

**Abstract**

The discovery of graphene and other two-dimensional (2-D) materials has stimulated a general interest in low-dimensional (low-D) materials. Whereas long time ago, Peierls (Peierls, 1935) and Landau's (Landau, 1937) theoretical work demonstrated that any one- and two-dimensional materials could not exist in any finite temperature environment. Then, two basic issues became a central concern for many researchers: How can stable low-D materials exist? What kind of low-D materials are stable? Here, we establish an energy stability criterion for low-D materials, which seeks to provide a clear answer to these questions. For a certain kind of element, the stability of its specific low-D structure is determined by several derivatives of its interatomic potential. This atomistic-based approach is then applied to study any straight/planar, low-D, equal-bond-length elemental materials. We found that 1-D monatomic chains, 2-D honeycomb lattices, square lattices, and triangular lattices are the only four permissible structures, and the stability of these structures can only be understood by assuming multi-body interatomic potentials. Using this approach, the stable existence of graphene, silicene and germanene can be explained.

*Keywords*: Existence criterion; Low-dimensional system; Interatomic potential; Mechanical stability

# 1. Introduction

Currently, one-dimensional (1-D) and two-dimensional (2-D) materials are very hot topics in research and development because of their extraordinary properties (Balandin and Nika, 2012; Casari et al., 2016; Geim and Novoselov, 2007; Novoselov, 2011). Scientists have successfully fabricated many types of the low-dimensional (low-D) materials, such as carbon atomic chains (Jin et al., 2009a), graphene (Novoselov et al., 2004), silicene (Feng et al., 2012), germanene (Davila et al., 2014) and so on (Balendhran et al., 2013; Ji et al., 2015; Jin et al., 2009b; Lu et al., 2014; Ohnishi et al., 1998; Pacile et al., 2008; Zhu et al., 2015). Peierls (Peierls, 1935) and Landau (Landau, 1937) proved some time ago that strictly 1-D and 2-D materials were thermodynamically unstable because of the infinite fluctuation of atomic displacement at any finite temperature, and their conclusion was later supported by Mermin and Wagner (Mermin, 1968; Mermin and Wagner, 1966). Using Bogoliubov's inequality, they demonstrated that the crystalline order could not be maintained in low-D materials (Mermin, 1968). Such an obvious contradiction between the aforementioned theoretical predictions (hereafter referred to as PLM theory) and the experimental results created much confusion in the scientific circles. Actually, the dilemma can be clarified as follows:

1. The atomic movements of low-D materials considered in PLM theory is constrained in the low-D space where they are defined. However, the low-D materials are embedded in the 3-D space. In fact, the bending vibration plays a significant role in the stability of low-D materials. Therefore, the PLM theory is not appropriate for suspended low-D materials.

2. According to PLM theory, the mean square fluctuation displacement of atoms $\overline{\mathbf{u}^2}$ in low-D materials becomes infinite when the dimensions of the body become arbitrarily large, and thus the crystalline orders of such materials are destroyed by the thermal fluctuations. However, in small-size, low-D materials, $\overline{\mathbf{u}^2}$ remains finite based on Peierls and Landau's derivation. As a result, it is not surprising that the finite-size, low-D materials can stably exist in a laboratory environment. Even the size of low-D materials becomes infinite, the fluctuation of the distance between neighboring atoms, such as atom 1 and atom 2 ($\overline{\left[\mathbf{u}(\mathbf{r_1}) - \mathbf{u}(\mathbf{r_2})\right]^2}$), is finite. This is to say, although $\overline{\mathbf{u}^2}$ is divergent, the chemical bonds ruptures would not occur. The structure can still maintained integrity in low-D lattices. Therefore, large low-D materials can stably exist.

Therefore, it can be found that the PLM theory cannot provide a solid criterion for the existence of low-D materials. Extensive research has been done on the stability of low-D materials (Fasolino et al., 2007; Los et al., 2015; O'hare et al., 2012; Zakharchenko et al., 2011), yet all of these works focus on the phase transition behaviors of these materials. Nonetheless, a criterion for the stable existence of low-D materials is still not available. To be more specific, it is still unclear what parameters determine the existence of low-D materials and what kind of element can form low-D materials.

The existence criterion for 3-D materials was worked out by Born and his co-workers (Born, 1940). In their works, the criterion was constructed from the positive definiteness of the atomistic-based macroscopic deformation energy based on the minimum free energy axiom, and it was expressed in the form of inequalities for the interatomic potential-related constants. Once the details of interatomic potential are known, we can determine whether the atoms can form a stable specific 3-D structure. In this paper, we develop an analytical atomistic-based energy approach to establish the existence criterion of low-D materials by considering their exotic low-D geometrical features. This approach relies on a two-step-deduction: deriving, first, the extreme value condition, and then, the positive definiteness of interatomic potential, and the criterion is expressed in the form of inequalities for several parameters of interatomic potential. For a brief explanation, we apply this general approach to derive the exact conditions determining the existence of straight/planar, low-D, equal-bond-length elemental materials for an arbitrary interatomic potential, and then the parameter-based phase diagrams are established to show the consequence of these existence criteria. The existence criteria of straight/planar low-D equal-bond-length elemental materials described by two typical types of atomistic potentials, pair potential and reactive empirical bond-order (REBO) potential (Brenner et al., 1991; Brenner et al., 2002; Stuart et al., 2000), are established. We find that the planar 2-D honeycomb structure is found to be the most stable one among all planar equal-bond-length structures, and the carbon has a stable planar 2-D honeycomb structure while silicon and germanium do not. Finally, the existence of the 2-D honeycomb-like buckled structure is also explored. It is shown that the honeycomb-like buckled structures of silicon and germanium are stable, which agrees with the theoretical and experimental studies of silicene and germanene (Cahangirov et al., 2009; Davila et al., 2014; Feng et al., 2012; O'hare et al., 2012).

## 2. An atomistic-based existence criterion for low-D materials

According to thermodynamics, it is well known that the free energy $F$ of a material should be kept to a minimum for its structure to be stable. The free energy of a crystal at finite temperature can be expressed in the form of:

$$F = \Phi - kT \ln Z_v, \tag{1}$$

where $\Phi$ is the total interatomic potential of the crystal, and $Z_v$ is the part of the partition function depending on the vibrations. However, it is difficult to apply Eq.(1) directly, because the complete phonon spectrum is difficult to derive analytically. Born (Born, 1940) proposed that the stability considerations remained valid when the thermodynamical system degenerated into a mechanical one. Here, analogous to Born's works, we assume that the temperature is sufficiently low, so the effect of vibrations can be neglected. The free energy $F$ becomes identical with the total interatomic potential $\Phi$. The existence criterion of low-D materials is established by the following two steps:

First, we determine the low-D structures that are permissible by deriving the extreme value condition of the potential energy of the system. Without loss of generality, the total interatomic potential of the materials have the form (Brenner et al., 1991; Brenner et al., 2002; Lennard-Jones and Hall, 1924; Los et al., 2005; Morse, 1929; Stuart et al., 2000):

$$\Phi = \frac{1}{2}\sum_{i}^{N}\phi_i\left(\{r_b\},\{cos\theta\},\{cos\varphi\}\right),$$

$$\phi_i = \sum_{l=1}^{m}\sum_{l}\phi_i^l\left(r_l,\{cos\theta\},\{cos\varphi\}\right) = \sum_{S}\phi_i^S\left(r_S,\{cos\theta\},\{cos\varphi\}\right) + \sum_{L=2}^{m}\sum_{L}\phi_i^L\left(r_L\right), \quad (2)$$

where $N$ and $\phi_i$ denote the number of atoms and double the potential energy of an atom $i$, respectively. $\{r_b\}$, $\{cos\theta\}$, and $\{cos\varphi\}$ are the sets of bond lengths, related bond angles and related torsion angles (the dihedral angle of two different atomic planes). $\phi_i^l$ describes the $l$th-nearest neighbor interactions, and $\phi_i^S$ describes the short-range interaction (nearest neighbor interaction), while $\phi_i^L$ is a pair potential that describes the long-range interaction ($L$th-nearest neighbor interactions). The variable $r_l$ is the $l$th-nearest neighbor atomic distance. $\sum_{l}$ denotes the summation over all the $l$th-nearest neighbor atoms. The upper bound of summation in $\sum_{l=1}^{m}$ means $m$th-nearest neighbor atomic distance is closest to the cutoff distance of interatomic potential.

The extreme value condition of $\Phi$ provides a set of equations on the permissible initial structures of materials at the equilibrium state. Generally, these equations are derived from $\left(\frac{\partial\Phi}{\partial r_b}\right)_0 = 0$, $\left(\frac{\partial\Phi}{\partial\theta}\right)_0 = const$ and $\left(\frac{\partial\Phi}{\partial\varphi}\right)_0 = const$, where the subscript "0" denotes the values at the equilibrium state. The latter two equations are usually obtained by solving a conditional extreme problem because the angles $\theta$ and $\varphi$ are not independent of each other. The equilibrium bond length $r_0$ can be derived by solving these equations.

Second, we determine the mechanical stability of the permissible low-D structure. For 3-D materials, the strain tensor $\{\varepsilon\}$ characterizes the macroscopic perturbation, thus the interatomic potential has the form $\Phi = \Phi(\{\varepsilon\})$. Then the mechanical stability conditions of materials can be derived from the positive definiteness of the quadratic terms of $\Phi$ with respect to $\{\varepsilon\}$. Different from 3-D materials, low-D materials are subjected to in-line/plane strain $\{\varepsilon\}$ and curvature $\{\kappa\}$, which cause by stretching and bending, respectively. $\{\varepsilon\} = \{\varepsilon_{xx}, \varepsilon_{yy}, \varepsilon_{xy}\}$ and $\{\kappa\} = \{\kappa_{xx}, \kappa_{yy}, \kappa_{xy}\}$ are tensors in 2-D materials, and they degenerate into scalars in 1-D materials. Then, interatomic potential can be expressed as a function of $\{\varepsilon\}$ and $\{\kappa\}$, i.e., $\Phi = \Phi(\{\varepsilon\}, \{\kappa\})$. The mechanical stability conditions of low-D materials can be derived from the positive definiteness of the quadratic terms of $\Phi$ with respect to $\{\varepsilon\}$ and $\{\kappa\}$. These terms have the form:

$$\Phi_{e_{\alpha\beta}, e_{\gamma\delta}} = \left(\frac{\partial^2 \Phi}{\partial e_{\alpha\beta} \partial e_{\gamma\delta}}\right)_0, \tag{3}$$

where $e_{\alpha\beta}, e_{\gamma\delta}$ denote the elements of $\{\varepsilon\}$ or $\{\kappa\}$.

According to Eq. (2), Eq. (3) can be expanded as follows:

$$\begin{aligned}
\Phi_{e_{\alpha\beta}, e_{\gamma\delta}} = \frac{1}{2}[&\sum_{l=1}^{m}\sum_{l}\frac{\partial \phi_i^l}{\partial r_l}\frac{\partial^2 r_l}{\partial e_{\alpha\beta} \partial e_{\gamma\delta}} + \sum_{S,\theta}\frac{\partial \phi_i^S}{\partial \cos\theta}\frac{\partial^2 \cos\theta}{\partial e_{\alpha\beta} \partial e_{\gamma\delta}} + \sum_{S,\varphi}\frac{\partial \phi_i^S}{\partial \cos\varphi}\frac{\partial^2 \cos\varphi}{\partial e_{\alpha\beta} \partial e_{\gamma\delta}} + \sum_{l=1}^{m}\sum_{l}\frac{\partial^2 \phi_i^l}{\partial r_l^2}\frac{\partial r_l}{\partial e_{\alpha\beta}}\frac{\partial r_l}{\partial e_{\gamma\delta}} \\
&+ \sum_{S}\sum_{\theta,\theta',\theta\neq\theta'}\frac{\partial^2 \phi_i^S}{\partial \cos\theta \partial \cos\theta'}\frac{\partial \cos\theta}{\partial e_{\alpha\beta}}\frac{\partial \cos\theta'}{\partial e_{\gamma\delta}} + \sum_{S}\sum_{\varphi,\varphi',\varphi\neq\varphi'}\frac{\partial^2 \phi_i^S}{\partial \cos\varphi \partial \cos\varphi'}\frac{\partial \cos\varphi}{\partial e_{\alpha\beta}}\frac{\partial \cos\varphi'}{\partial e_{\gamma\delta}} \\
&+ \sum_{S,\theta}\frac{\partial^2 \phi_i^S}{\partial r_S \partial \cos\theta}\left(\frac{\partial r_S}{\partial e_{\alpha\beta}}\frac{\partial \cos\theta}{\partial e_{\gamma\delta}} + \frac{\partial r_S}{\partial e_{\gamma\delta}}\frac{\partial \cos\theta}{\partial e_{\alpha\beta}}\right) + \sum_{S,\varphi}\frac{\partial^2 \phi_i^S}{\partial r_S \partial \cos\varphi}\left(\frac{\partial r_S}{\partial e_{\alpha\beta}}\frac{\partial \cos\varphi}{\partial e_{\gamma\delta}} + \frac{\partial r_S}{\partial e_{\gamma\delta}}\frac{\partial \cos\varphi}{\partial e_{\alpha\beta}}\right) \\
&+ \sum_{S,\theta,\varphi}\frac{\partial^2 \phi_i^S}{\partial \cos\theta \partial \cos\varphi}\left(\frac{\partial \cos\theta}{\partial e_{\alpha\beta}}\frac{\partial \cos\varphi}{\partial e_{\gamma\delta}} + \frac{\partial \cos\theta}{\partial e_{\gamma\delta}}\frac{\partial \cos\varphi}{\partial e_{\alpha\beta}}\right)]_0
\end{aligned}$$

(4)

Equation (4) expresses $\Phi_{e_{\alpha\beta}, e_{\gamma\delta}}$ by several derivatives of the potential (e.g., $\frac{\partial \phi_i^l}{\partial r_l}$, $\frac{\partial \phi_i^S}{\partial \cos \theta}$, $\frac{\partial \phi_i^S}{\partial \cos \varphi}$, $\frac{\partial^2 \phi_i^l}{\partial r_l^2}$ etc.) at the equilibrium state, and they depend on the analytical form of interatomic potential.

The coefficients of these derivatives of the potential (e.g., $\frac{\partial^2 \cos \theta}{\partial e_{\alpha\beta} \partial e_{\gamma\delta}}$, $\frac{\partial^2 \cos \varphi}{\partial e_{\alpha\beta} \partial e_{\gamma\delta}}$, $\frac{\partial r_l}{\partial e_{\alpha\beta}} \frac{\partial r_l}{\partial e_{\gamma\delta}}$ etc.) can be derived from the atomistic-based continuum theory (Arroyo and Belytschko, 2002; Born and Huang, 1998; Zhang et al., 2004), which links the macroscopic deformation of an atomistic system to that of a continuum. The Cauchy–Born rule (Born and Huang, 1998) describes the homogenous deformation of 3-D materials, and the distance between two atoms $i$ and $j$ is affected by strain tensor $\{\varepsilon\}$: $|\mathbf{r}_{ij}| = r_{ij}^0 \sqrt{(\delta_{\alpha\beta} + 2\varepsilon_{\alpha\beta})(n_\alpha + x_\alpha)(n_\beta + x_\beta)}$, where $r_{ij}^0$, $n_\alpha$ and $n_\beta$ are the length and direction vectors between atoms $i$ and $j$ before deformation. $x_\alpha$ and $x_\beta$ are components of the shift vector, and they are nonzero when atoms $i$ and $j$ are located in different sub-lattices, $\alpha, \beta = x, y, z$. Consider a strictly low-D material (Novoselov et al., 2005) (i.e., it could be regarded as a single or several smoothly curved lines/surfaces). When the initial radius of curvature is much larger than the atomic spacing, the vector between two atoms $i$ and $j$ on the same line/surface for infinitesimal curvature $\{\kappa\}$ can be represented by (Wu et al., 2008)

$$\mathbf{r}_{ij} \approx \left[ r_{ij}^0 (n_\alpha + x_\alpha) - \frac{1}{6} \kappa_{\beta\lambda} \kappa_{\gamma\mu} T^{\alpha\mu} (n_\beta + x_\beta)(n_\gamma + x_\gamma)(n_\mu + x_\mu)(r_{ij}^0)^3 \right] \mathbf{T}_\alpha \qquad (5)$$
$$+ \frac{1}{2} \kappa_{\alpha\beta} (n_\alpha + x_\alpha)(n_\beta + x_\beta)(r_{ij}^0)^2 \mathbf{N},$$

where $\mathbf{T}_\alpha$ are the covariant base vectors lying on the tangent line/surface. $T^{\alpha\mu}$ are the contravariant components of metric tensors for the covariant base vectors. $\mathbf{N}$ is the unit normal vector to the line/surface. In addition, $\alpha, \beta, \gamma, \mu = x$ for 1-D materials and $\alpha, \beta, \gamma, \mu = x, y$ for 2-D materials.

Note that $\mathbf{T}_\alpha \cdot \mathbf{T}_\beta = \delta_{\alpha\beta} + 2\varepsilon_{\alpha\beta}$, and it could be found that (Huang et al., 2006):

$$\left|\mathbf{r}_{ij}\right| \approx r_{ij}^0 \sqrt{\left(\delta_{\alpha\beta} + 2\varepsilon_{\alpha\beta}\right)\left(n_\alpha + x_\alpha\right)\left(n_\beta + x_\beta\right) - \frac{[\kappa_{\alpha\beta}(n_\alpha + x_\alpha)(n_\beta + x_\beta)]^2}{12}\left(r_{ij}^0\right)^2}. \quad (6)$$

where $\alpha, \beta = x$ for 1-D materials and $\alpha, \beta = x, y$ for 2-D materials. Equation (6) reduces to the Cauchy–Born rule when $\kappa_{\alpha\beta} = 0$.

$\theta_{ijk}$ and $\varphi_{ijkl}$ can respectively be expressed as:

$$cos\theta_{ijk} = \frac{\mathbf{r}_{ij} \, \mathbf{r}_{ik}}{\left|\mathbf{r}_{ij}\right|\left|\mathbf{r}_{ik}\right|}, \quad (7)$$

$$cos\varphi_{ijkl} = \mathbf{N}_{ijk} \, \mathbf{N}_{ijl} = \frac{\mathbf{r}_{ik} \times \mathbf{r}_{ij}}{\left|\mathbf{r}_{ik} \times \mathbf{r}_{ij}\right|} \frac{\mathbf{r}_{ij} \times \mathbf{r}_{il}}{\left|\mathbf{r}_{ij} \times \mathbf{r}_{il}\right|}, \quad (8)$$

where $\mathbf{N}_{ijk}$ and $\mathbf{N}_{ijl}$ are the unit normal vectors of atomic plane $ijk$ and atomic plane $ijl$.

The shift vector $\mathbf{x}$ is determined by minimizing the deform potential energy with respect to $\mathbf{x}$:

$$\frac{\partial \Phi}{\partial \mathbf{x}} = 0, \quad (9a)$$

which gives:

$$\mathbf{x} = \mathbf{x}\left(\{\varepsilon\}, \{\kappa\}\right). \quad (9b)$$

In the case of strictly low-D materials, the coefficients of derivatives of the potential in Eq. (4) can be obtained from Eqs. (6)–(9). These coefficients are determined by the structure-related parameters.

The criterion constructed by the two-step deduction provides basic conditions for any low-D material to exist. Since different elements possesses different interatomic potential, one can clarify what kinds of elements will form stable low-D materials according to this criterion.

## 3. Existence criterion for straight/planar, low-D, equal-bond-length elemental materials

In this section, we apply the general methodology introduced above to derive the exact formula determining the existence of low-D materials with a specific structure. Generally speaking, it must be noted that not all bonds in low-D materials are equal in length, e.g., carbyne, graphdiyne and graphyne (Ivanovskii, 2013). For simplicity, we restrict our consideration to straight/planar, low-D elemental materials where all bond lengths are equal. We assume that the sizes of these systems are quite large and the boundary effect can be eliminated.

### 3.1. Permissible structures

First of all, we analyze the permissible structures of straight 1-D and planar 2-D equal-bond-length elemental materials, respectively.

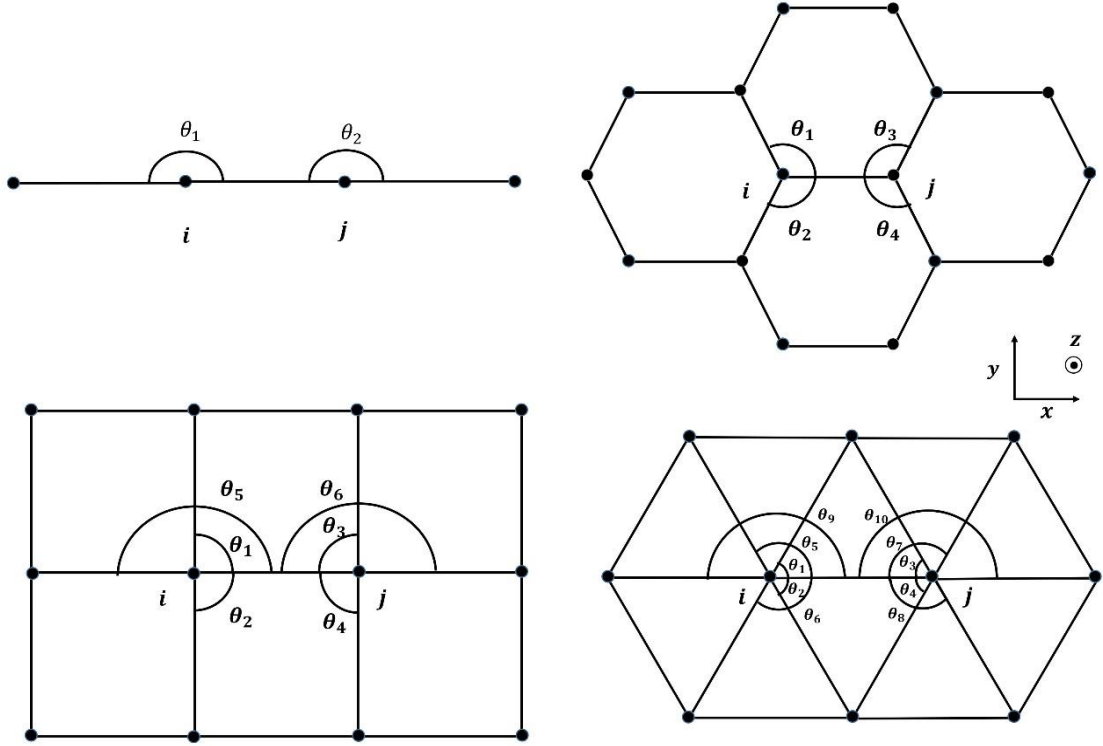

Figure 1. The permissible structures of straight 1-D and planar 2-D equal-bond-length elemental materials. (a) 1-D monatomic chain. (b) Planar 2-D Honeycomb structure. (c) Planar 2-D square structure. (d) Planar 2-D triangular structure.

*3.1.1. Straight 1-D equal-bond-length elemental materials*

The straight 1-D equal-bond-length elemental materials are monatomic chains. For the 1-D atomic chain, assume that every atom is equal in energy, the total interatomic potential has the form:

$$\Phi = \frac{1}{2} N \sum_{l=1}^{,m} \sum_{l} \phi_i^l = \frac{1}{2} N \left[ \sum_{S} \phi_i^S \left( r_S, \{cos\theta\} \right) + \sum_{L=2}^{,m} \sum_{L} \phi_i^L \left( r_L \right) \right]. \quad (10)$$

One should notice that $|r_l| = l r_b$. The related bond angles are shown in Fig. 1(a).

Since there is no geometric constraint for the arguments of $\Phi$, the requirements of extreme value condition for the potential energy are:

$$\left(\frac{\partial \Phi}{\partial r_b}\right)_0 = 0, \tag{11a}$$

$$\left(\frac{\partial \Phi}{\partial \theta}\right)_0 = \left(\frac{\partial \Phi}{\partial \cos\theta}\right)_0 (-\sin\theta)_{\theta=\pi} = 0. \tag{11b}$$

According to Eq. (11b), since $(-\sin\theta)_{\theta=\pi} = 0$, $\left(\frac{\partial \Phi}{\partial \cos\theta}\right)_0$ is probably nonzero and the first derivative terms of $\Phi$ with respect to $\cos\theta$ in Eq. (4) do not vanish.

*3.1.2. Planar 2-D equal-bond-length elemental materials*

The interatomic potential of planar 2-D equal-bond-length elemental materials is shown in Eq. (2). In this case, the torsion angles are equal to zero in ground state. Since there is no geometric constraint for bond length and torsion angles in this case, it can be similarly found that:

$$\left(\frac{\partial \Phi}{\partial r_b}\right)_0 = 0, \tag{12a}$$

$$\left(\frac{\partial \Phi}{\partial \varphi}\right)_0 = \left(\frac{\partial \Phi}{\partial \cos\varphi}\right)_0 (-\sin\varphi)_{\varphi=0} = 0. \tag{12b}$$

Since $(-\sin\varphi)_{\varphi=0} = 0$, $\left(\frac{\partial \Phi}{\partial \cos\varphi}\right)_0$ is probably nonzero. The first derivative terms of $\Phi$ respect to $\cos\varphi$ in Eq. (4) do not vanish.

The total interatomic potential $\Phi$ takes the whole bond angle in the materials into account. We assume that there are M bonds connecting an atom $i$ with its neighboring atoms. Thus the number of the bond angles which atom $i$ is at the vertex of the angles is $M\left[\frac{M}{2}\right]$, where $[\ ]$ denotes the bracket function. These angles can be classified as $\left[\dfrac{M}{2}\right]$ sets ($\{\theta^1\},\{\theta^2\}...\{\theta^{\left[\frac{M}{2}\right]}\}$). The element $\theta_n^m$ of the set $\{\theta^m\}$ is the angle whose initial side is the *n*th bond and the terminal side is its *m*th neighboring bond (counter clockwise direction). As $\sum_{n=1}^{M}\theta_n^m = 2m\pi$ ( $m=1,2\ ...\ \left[\dfrac{M}{2}\right]$ ), several equations are obtained using the Lagrange conditional extremum theory:

$$\frac{\partial \Phi}{\partial \theta_1^1} = \frac{\partial \Phi}{\partial \theta_2^1} \ ... \ = \frac{\partial \Phi}{\partial \theta_M^1} = \lambda_1,$$

$$\frac{\partial \Phi}{\partial \theta_1^2} = \frac{\partial \Phi}{\partial \theta_2^2} \ ... \ = \frac{\partial \Phi}{\partial \theta_M^2} = \lambda_2,$$

$$...$$

$$\frac{\partial \Phi}{\partial \theta_1^{\left[\frac{M}{2}\right]}} = \frac{\partial \Phi}{\partial \theta_2^{\left[\frac{M}{2}\right]}} \ ... \ = \frac{\partial \Phi}{\partial \theta_M^{\left[\frac{M}{2}\right]}} = \lambda_{\left[\frac{M}{2}\right]},$$

(12c)

where $\lambda_1,\lambda_2\ ...\ \lambda_{\left[\frac{M}{2}\right]}$ are Lagrange multipliers and they are probably nonzero. As $\left(\dfrac{\partial \Phi}{\partial \cos\theta_n^m}\right)_0 = -\left(\dfrac{1}{\sin\theta_n^m}\dfrac{\partial \Phi}{\partial \theta_n^m}\right)_0 = -\dfrac{\lambda_m}{\left(\sin\theta_n^m\right)_0}$, the first derivative terms of $\Phi$ respect to $\cos\theta$ in Eq. (4) do not vanish.

For the concrete form of multi-body potential, we assume $\dfrac{\partial \Phi}{\partial \theta}$ varies monotonically with $\theta$ change. Thus, Eq. (12c) implies that all the elements of a set $\{\theta^m\}$ are equal. Remembering that translational symmetry is present in a crystal lattice, we draw a conclusion that the planar 2-D equal-bond-length elemental materials have three permissible initial configurations: honeycomb structure, square structure and triangular structure.

### 3.2. Mechanical stability

As mentioned above, the mechanical stability conditions of low-D materials can be derived from the positive definiteness of the quadratic terms of $\Phi$ with respect to $\{\varepsilon\}$ and $\{\kappa\}$. Since $\{\varepsilon\}$ and $\{\kappa\}$ are independent of one another geometrically in this case, strain and curvature are decoupled in energy. Therefore, $\Phi_{\{\varepsilon\},\{\kappa\}}=0$, and the remainder of quadratic terms form a matrix: $[\mathbf{M}]=N\begin{bmatrix}[C] & \mathbf{0} \\ \mathbf{0} & [D]\end{bmatrix}$, where the elements of $[C]$ and $[D]$ are $\frac{1}{N}\Phi_{\{\varepsilon\},\{\varepsilon\}}$ and $\frac{1}{N}\Phi_{\{\kappa\},\{\kappa\}}$, respectively. In this sense, the positive definiteness of $[C]$ and $[D]$ reflect the need for stability of low-D materials.

It is noteworthy that $[D]$ shows the leading feature of stability of low-D materials. Unlike 3-D materials, the low-D materials are extremely anisotropic because they lack restriction in the out-of-line/plane direction. Thus, the out-of-line/plane perturbation has a much greater influence on the stability of low-D materials than the in-line/plane perturbation. Because the positive definiteness of $[C]$ was introduced in the search for in-line/plane stability conditions completed by Born, here, we present $[D]$, which is only present in low-D materials, to clarify the out-of-line/plane stability which is not a concerned in PLM theory.

### 3.2.1. Straight 1-D equal-bond-length elemental materials

Since 1-D monatomic chains possess simple lattice structure, the shift vector in Eq. (9) vanishes. Besides, the in-line strain $\{\varepsilon\}$ and curvature $\{\kappa\}$ are scalars in this case. Thus, Eqs. (6) and (7) have the forms:

$$r_{ij} = r_{ij}^{0}\sqrt{1+2\varepsilon-\frac{1}{12}\kappa^{2}\left(r_{ij}^{0}\right)^{2}}, \tag{13}$$

$$\cos\theta_{ijk} = \frac{r_0^2}{r_{ij}r_{ik}}\left[-(1+2\varepsilon)+\frac{7}{12}\kappa^2 r_0^2\right], \quad (14)$$

where $r_0$ is equilibrium bond length, which is obtained by solving Eq. (11a). Furthermore, $[C]$ and $[D]$ are degenerated into scalars:

$$\begin{aligned} C &= \frac{1}{N}\Phi_{\varepsilon,\varepsilon}, \\ D &= \frac{1}{N}\Phi_{\kappa,\kappa}. \end{aligned} \quad (15)$$

In substituting Eqs. (10), (15), and (16) into Eq. (4), the accurate expressions of $C$ and $D$ are:

$$\begin{aligned} C &= \frac{1}{2}\left[\sum_{l=1}^{m}\sum_{l}\left(\frac{\partial^2\phi_i^l}{\partial r_l^2}\right)r_l^2\right]_0, \\ D &= 2\left(\frac{\partial \phi_i^S}{\partial \cos\theta}\right)_{\theta=\pi} r_0^2 - \frac{1}{24}\left[\sum_{l=1}^{m}\sum_{l}\left(\frac{\partial \phi_i^l}{\partial r_l}\right)r_l^3\right]_0. \end{aligned} \quad (16)$$

Note that $\left[\sum_{l=1}^{m}\sum_{l}\left(\frac{\partial^2\phi_i^l}{\partial r_l^2}\right)r_l^2\right]_0 = \left(\frac{\partial^2\phi_i}{\partial r_b^2}\right)_0 r_0^2$, and the mechanical stability condition of 1-D materials becomes:

$$\left(\frac{\partial^2\phi_i}{\partial r_b^2}\right)_0 > 0, \quad (17a)$$

$$\left(\frac{\partial \phi_i^S}{\partial \cos\theta}\right)_{\theta=\pi} r_0^2 - \frac{1}{48}\left[\sum_{l=1}^{m}\sum_{l}\left(\frac{\partial \phi_i^l}{\partial r_l}\right)r_l^3\right]_0 > 0. \quad (17b)$$

Equation (17a) guarantees that 1-D materials cannot break up along the axial direction spontaneously, and (17b) indicates that 1-D materials possess the ability to resisting out-of-line perturbation.

Take carbon atomic chains, for instance, according to second-generation REBO potential for carbon bonding (Brenner et al., 2002), we have:

$$\left(\frac{\partial^2 \phi_i}{\partial r_b^2}\right)_0 = 54.34 \, eV/\text{Å}^2 > 0,$$

$$\left(\frac{\partial \phi_i^S}{\partial \cos\theta}\right)_{\theta=\pi} r_0^2 - \frac{1}{48}\sum_{l=1}^m \sum_l \left(\frac{\partial \phi_i^l}{\partial r_l}\right) r_l^3 = 1.14 \, eV \cdot \text{Å}^2 > 0.$$

It implies that the carbon atomic chains could stably exist and possess the ability to resist in-line and out-of-line perturbation.

*3.2.2. Planar 2-D equal-bond-length elemental materials*

For planar 2-D equal-bond-length elemental materials, the two matrixes $[\mathbf{C}]$ and $[\mathbf{D}]$ are introduced as:

$$[\mathbf{C}] = \frac{1}{N} \begin{bmatrix} \Phi_{\varepsilon_{xx},\varepsilon_{xx}} & \Phi_{\varepsilon_{xx},\varepsilon_{yy}} & \Phi_{\varepsilon_{xx},\varepsilon_{xy}} \\ \Phi_{\varepsilon_{yy},\varepsilon_{xx}} & \Phi_{\varepsilon_{yy},\varepsilon_{yy}} & \Phi_{\varepsilon_{yy},\varepsilon_{xy}} \\ \Phi_{\varepsilon_{xy},\varepsilon_{xx}} & \Phi_{\varepsilon_{xy},\varepsilon_{yy}} & \Phi_{\varepsilon_{xy},\varepsilon_{xy}} \end{bmatrix}, \tag{18a}$$

$$[\mathbf{D}] = \frac{1}{N} \begin{bmatrix} \Phi_{\kappa_{xx},\kappa_{xx}} & \Phi_{\kappa_{xx},\kappa_{yy}} & \Phi_{\kappa_{xx},\kappa_{xy}} \\ \Phi_{\kappa_{yy},\kappa_{xx}} & \Phi_{\kappa_{yy},\kappa_{yy}} & \Phi_{\kappa_{yy},\kappa_{xy}} \\ \Phi_{\kappa_{xy},\kappa_{xx}} & \Phi_{\kappa_{xy},\kappa_{yy}} & \Phi_{\kappa_{xy},\kappa_{xy}} \end{bmatrix}. \tag{18b}$$

The mechanical stability conditions can be derived from the positive definiteness of $[\mathbf{C}]$ and $[\mathbf{D}]$, and they correspond to in-plane stability and out-of-plane stability, respectively.

Here, we discuss the mechanical stability conditions of planar 2-D elemental materials with the three permissible structures: honeycomb structure, square structure and triangular structure, respectively.

The honeycomb structure is a compound structure which is formed by two triangular sub-lattices. Figure 1(b) illustrates the Honeycomb structure, with the related angles of interaction between atom $i$ and atom $j$.

Equation (2) shows the general form of total interatomic potential. Considering the symmetry in honeycomb structure, according to Eqs.(4)–(9) and (20), we have:

$$[\mathbf{C}] = \begin{bmatrix} C_{11} & C_{12} & 0 \\ C_{21} & C_{22} & 0 \\ 0 & 0 & C_{33} \end{bmatrix}, \quad (19a)$$

$$[\mathbf{D}] = \begin{bmatrix} D_{11} & D_{12} & 0 \\ D_{21} & D_{22} & 0 \\ 0 & 0 & D_{33} \end{bmatrix}, \quad (19b)$$

where

$$\begin{aligned} C_{11} &= C_{22} = \frac{1}{2}\left[\sum_{l=1}^{m}\sum_{l}\left(\frac{\partial^2 \phi_i^l}{\partial r_l^2}\right) r_l^2 \left(n_x^l\right)^4 + \frac{B}{16}\right]_0, \\ C_{12} &= C_{21} = \frac{1}{2}\left[\sum_{l=1}^{m}\sum_{l}\left(\frac{\partial^2 \phi_i^l}{\partial r_l^2}\right) r_l^2 \left(n_x^l\right)^2 \left(n_y^l\right)^2 - \frac{B}{16}\right]_0, \\ C_{33} &= 2\left[\sum_{l=1}^{m}\sum_{l}\left(\frac{\partial^2 \phi_i^l}{\partial r_l^2}\right) r_l^2 \left(n_x^l\right)^2 \left(n_y^l\right)^2 + \frac{B}{16}\right]_0, \end{aligned} \quad (20a)$$

and

$$\begin{aligned} D_{11} &= D_{22} = \frac{3}{8} r_0^2 \left[3\left(\frac{\partial \phi_i^S}{\partial \cos\theta}\right)_{\theta=\frac{2\pi}{3}} - 14\left(\frac{\partial \phi_i^S}{\partial \cos\varphi}\right)_{\varphi=0}\right] - \frac{1}{24}\left[\sum_{l=1}^{m}\sum_{l}\left(\frac{\partial \phi_i^l}{\partial r_l}\right) r_l^3 \left(n_x^l\right)^4\right]_0, \\ D_{12} &= D_{21} = \frac{3}{8} r_0^2 \left[3\left(\frac{\partial \phi_i^S}{\partial \cos\theta}\right)_{\theta=\frac{2\pi}{3}} + 2\left(\frac{\partial \phi_i^S}{\partial \cos\varphi}\right)_{\varphi=0}\right] - \frac{1}{24}\left[\sum_{l=1}^{m}\sum_{l}\left(\frac{\partial \phi_i^l}{\partial r_l}\right) r_l^3 \left(n_x^l\right)^2 \left(n_y^l\right)^2\right]_0, \\ D_{33} &= -12 r_0^2 \left(\frac{\partial \phi_i^S}{\partial \cos\varphi}\right)_{\varphi=0} - \frac{1}{6}\left[\sum_{l=1}^{m}\sum_{l}\left(\frac{\partial \phi_i^l}{\partial r_l}\right) r_l^3 \left(n_x^l\right)^2 \left(n_y^l\right)^2\right]_0. \end{aligned} \quad (20b)$$

The coefficient $B$ in Eq. (20a) is:

$$\begin{aligned} B &= 9(1-A)^2 \left[4\left(\frac{\partial \phi_i^S}{\partial \cos\theta}\right)_{\theta=\frac{2\pi}{3}} + 6\left(\frac{\partial^2 \phi_i^S}{\partial \cos\theta^2}\right)_{\theta=\frac{2\pi}{3}} - 3\left(\frac{\partial^2 \phi_i^S}{\partial \cos\theta \partial \cos\theta'}\right)_{\theta,\theta'=\frac{2\pi}{3}}\right] - 36 r_0 (1-A^2)\left(\frac{\partial^2 \phi_i^S}{\partial r_s \partial \cos\theta}\right)_{\theta=\frac{2\pi}{3}} \\ &+ 4\left[A^2 \sum_{l=odd}^{m}\sum_{l}\left(\frac{\partial \phi_i^l}{\partial r_l} + \frac{\partial^2 \phi_i^l}{\partial r_l^2} r_l\right) r_l \left(n_x^l\right)^2\right]_0 - 16\left[A\sum_{l=odd}^{m}\sum_{l}\left(\frac{\partial \phi_i^l}{\partial r_l} - \frac{\partial^2 \phi_i^l}{\partial r_l^2} r_l\right) r_l \left(n_x^l\right)^3\right]_0 \end{aligned} \quad (21a)$$

where

$$A = \frac{36\left(\frac{\partial \phi_i^S}{\partial \cos\theta}\right)_{\theta=\frac{2\pi}{3}} + 54\left(\frac{\partial^2 \phi_i^S}{\partial \cos\theta^2}\right)_{\theta=\frac{2\pi}{3}} - 27\left(\frac{\partial^2 \phi_i^S}{\partial \cos\theta \partial \cos\theta'}\right)_{\theta,\theta'=\frac{2\pi}{3}} + 8\sum_{l=odd}^{m}\sum_{l}\left(\frac{\partial \phi_i^l}{\partial r_l} - \frac{\partial^2 \phi_i^l}{\partial r_l^2}r_l\right)_0 r_l (n_x^l)^3}{36\left(\frac{\partial \phi_i^S}{\partial \cos\theta}\right)_{\theta=\frac{2\pi}{3}} + 54\left(\frac{\partial^2 \phi_i^S}{\partial \cos\theta^2}\right)_{\theta=\frac{2\pi}{3}} + 36r_0\left(\frac{\partial^2 \phi_i^S}{\partial r_S \partial \cos\theta}\right)_{\theta=\frac{2\pi}{3}} - 27\left(\frac{\partial^2 \phi_i^S}{\partial \cos\theta \partial \cos\theta'}\right)_{\theta,\theta'=\frac{2\pi}{3}} + 4\sum_{l=odd}^{m}\sum_{l}\left(\frac{\partial^2 \phi_i^l}{\partial r_l^2}\right)_0 r_l^2 (n_x^l)^2}. \quad (21b)$$

Here, $r_0$ is equilibrium bond length, and $n_x^l$ and $n_y^l$ are the components of the $l$th neighbor direction vector before deformation.

The positive definiteness of $[\mathbf{C}]$ obeys the requirements:

$$\begin{aligned} C_{11} + C_{12} &> 0 \\ C_{11} - C_{12} &> 0. \\ C_{33} &> 0 \end{aligned} \quad (22a)$$

The positive definiteness of $[\mathbf{D}]$ obeys the requirements:

$$\begin{aligned} D_{11} + D_{12} &> 0 \\ D_{11} - D_{12} &> 0. \\ D_{33} &> 0 \end{aligned} \quad (22b)$$

If the conditions in Eq. (22a) are not satisfied, the honeycomb structure cannot resist the in-plane distortion. When $C_{11} + C_{12} \leq 0$, the atoms in the materials possess no cohesion and these atoms cannot form a solid or even a fluid. When $C_{11} - C_{12} \leq 0$, the equibiaxial tension-compression happens spontaneously, and the materials lose the ability to resist uniaxial compression. When $C_{33} \leq 0$, the ability to resist shear deformation vanishes, and it manifests as a fluid. Thus, if the condition in Eq. (22a) is not satisfied, the material can be considered to be a gas, gel, or fluid, rather than a solid. This is in accord with Born's results (Born, 1939), which are obtained from an investigation of the stability of 3-D materials.

Moreover, if conditions in Eq. (22b) are violated, the honeycomb structure cannot resist out-of-plane perturbation. When $D_{11} + D_{12} \leq 0$ or $D_{11} - D_{12} \leq 0$, the 2-D material becomes crimped or saddle-like spontaneously. When $D_{33} \leq 0$, the spontaneous twist of suspended 2-D materials occurs. Thus, if the condition in (22b) is not satisfied, the honeycomb configuration of 2-D materials cannot stably exist. It may change to other configuration of low-D materials or quasi low-D materials, or even 3-D materials.

Generally speaking, the long-range interactions are weak, especially for covalent system (e.g., the cutoff radius for carbon bonding is $2.0\,\text{Å}$ (Brenner et al., 2002)), and it can be assumed that the interatomic potential only possesses a short-range interaction, that is to say, $\phi_i^L = 0$, $\left(\dfrac{\partial \phi_i^S}{\partial r_S}\right)_0 = 0$. The total interatomic potential becomes:

$$\Phi = \frac{1}{2} N \sum_S \phi_i^S \left(r_S, \{cos\theta\}, \{cos\varphi\}\right). \tag{23}$$

Then we have:

$$C_{11} = C_{22} = \frac{9}{16} r_0^2 \left(\frac{\partial^2 \phi_i^S}{\partial r_S^2}\right)_0 + \frac{1}{32} B$$

$$C_{12} = C_{21} = \frac{3}{16} r_0^2 \left(\frac{\partial^2 \phi_i^S}{\partial r_S^2}\right)_0 - \frac{1}{32} B, \tag{24a}$$

$$C_{33} = \frac{3}{4} r_0^2 \left(\frac{\partial^2 \phi_i^S}{\partial r_S^2}\right)_0 + \frac{1}{8} B$$

and

$$D_{11} = D_{22} = \frac{3}{8} r_0^2 \left[ 3 \left(\frac{\partial \phi_i^S}{\partial \cos\theta}\right)_{\theta=\frac{2\pi}{3}} - 14 \left(\frac{\partial \phi_i^S}{\partial \cos\varphi}\right)_{\varphi=0} \right]$$

$$D_{12} = D_{21} = \frac{3}{8} r_0^2 \left[ 3 \left(\frac{\partial \phi_i^S}{\partial \cos\theta}\right)_{\theta=\frac{2\pi}{3}} + 2 \left(\frac{\partial \phi_i^S}{\partial \cos\varphi}\right)_{\varphi=0} \right]. \tag{24b}$$

$$D_{33} = -12 r_0^2 \left(\frac{\partial \phi_i^S}{\partial \cos\varphi}\right)_{\varphi=0}$$

The mechanical stability conditions become

$$\begin{aligned} C_{11}+C_{12} &> 0 \\ C_{11}-C_{12} &> 0 \\ C_{33} &> 0 \end{aligned} \rightarrow \begin{aligned} \left(\frac{\partial^2 \phi_i^S}{\partial r_S^2}\right)_0 &> 0 \\ 6r_0^2\left(\frac{\partial^2 \phi_i^S}{\partial r_S^2}\right)_0 + B &> 0 \end{aligned} ,\qquad (25a)$$

and

$$\begin{aligned} D_{11}+D_{12} &> 0 \\ D_{11}-D_{12} &> 0 \\ D_{33} &> 0 \end{aligned} \rightarrow \begin{aligned} \left(\frac{\partial \phi_i^S}{\partial \cos\theta}\right)_{\theta=\frac{2\pi}{3}} - 2\left(\frac{\partial \phi_i^S}{\partial \cos\varphi}\right)_{\varphi=0} &> 0 \\ \left(\frac{\partial \phi_i^S}{\partial \cos\varphi}\right)_{\varphi=0} &< 0 \end{aligned} .\qquad (25b)$$

Graphene is a typical 2-D elemental material with planar honeycomb structure, which is made up of carbon atoms. According to second-generation REBO potential for carbon bonding (Brenner et al., 2002), we have:

$$\begin{aligned} \left(\frac{\partial^2 \phi_i^S}{\partial r_S^2}\right)_0 &= 43.56\,eV\!/\!\text{Å}^2 > 0 \\ 6r_0^2\left(\frac{\partial^2 \phi_i^S}{\partial r_S^2}\right)_0 + B &= 455.05\,eV > 0 \\ \left(\frac{\partial \phi_i^S}{\partial \cos\theta}\right)_{\theta=\frac{2\pi}{3}} - 2\left(\frac{\partial \phi_i^S}{\partial \cos\varphi}\right)_{\varphi=0} &= 1.50\,eV > 0 \\ \left(\frac{\partial \phi_i^S}{\partial \cos\varphi}\right)_{\varphi=0} &= -0.35\,eV < 0 \end{aligned}.$$

As a result, carbon has a stable planar honeycomb structure.

Here, we clarify whether the other group-IV elements such as silicon and germanium have a stable planar honeycomb structure. Since the atoms of silicon and germanium are larger than that of carbon, the long-range interaction of these atoms also vanishes. We can adopt the Tersoff-type potentials for silicon (Erhart and Albe, 2005) and germanium(Tersoff, 1989) bonding. The conditions in (27a) for silicon are

$$\left(\frac{\partial^2 \phi_i^S}{\partial r_{ij}^2}\right)_0 = 10.70\,eV/\text{Å}^2 > 0$$

$$6r_0^2\left(\frac{\partial^2 \phi_i^S}{\partial r_{ij}^2}\right)_0 + B = 180.98\,eV > 0$$

and for germanium are

$$\left(\frac{\partial^2 \phi_i^S}{\partial r_{ij}^2}\right)_0 = 8.26\,eV/\text{Å}^2 > 0$$

$$6r_0^2\left(\frac{\partial^2 \phi_i^S}{\partial r_{ij}^2}\right)_0 + B = 332.00\,eV > 0$$

As a result, the planar honeycomb structures of silicon and germanium are stable when the systems are subjected to in-plane perturbations. Whereas, due to weak $\pi$ bonding (Cahangirov et al., 2009), the torsion term in interatomic potentials for silicon and germanium is negligible. Therefore, $\left(\frac{\partial \phi_i^S}{\partial \cos\varphi}\right)_{\varphi=0}$ for silicon and germanium are approximately equal to zero, which may lead to non-positive definiteness of $[\mathbf{D}]$. For a Tersoff-type potential, the values of $\left(\frac{\partial \phi_i^S}{\partial \cos\theta}\right)_{\theta=\frac{2\pi}{3}} - 2\left(\frac{\partial \phi_i^S}{\partial \cos\varphi}\right)_{\varphi=0}$ are $-0.30\,eV$ for silicon and $-0.17\,eV$ for germanium, respectively. This means that the planar honeycomb structures of silicon and germanium cannot resist out-of-plane perturbation, and they are wrinkled spontaneously.

This conclusion was also obtained by first-principle studies (Cahangirov et al., 2009; Sahin et al., 2009). They reveal that the out-of-plane optical (ZO) phonon modes of planar honeycomb structures of silicon and germanium have imaginary frequencies (softened modes) and the mode at the $\Gamma$ point is the most softened. Therefore, planar honeycomb structures of silicon and germanium are unstable. Moreover, they find that honeycomb-like buckled structures are energetically favorable. By solving dynamics equations for lattice vibration, we found that the ZO mode at the $\Gamma$ point is proportional to $\left(\frac{\partial \phi_i^S}{\partial \cos\theta}\right)_{\theta=\frac{2\pi}{3}} - 2\left(\frac{\partial \phi_i^S}{\partial \cos\varphi}\right)_{\varphi=0}$ by only considering the short-range interaction. As mentioned above, we have proved that planar honeycomb structures of silicon and germanium are unstable because of negative $\left(\frac{\partial \phi_i^S}{\partial \cos\theta}\right)_{\theta=\frac{2\pi}{3}} - 2\left(\frac{\partial \phi_i^S}{\partial \cos\varphi}\right)_{\varphi=0}$. Therefore, this instability condition implies that long-wave ZO modes are softened, and these soft-modes lead to a spontaneously relative vertical displacement of alternate atoms. Then the planar honeycomb structure may be translated to a honeycomb-like buckled structure. This is consistent with the results of the aforementioned first-principles calculations.

The phase diagrams derived from Eq. (25) are shown in Fig. 2.

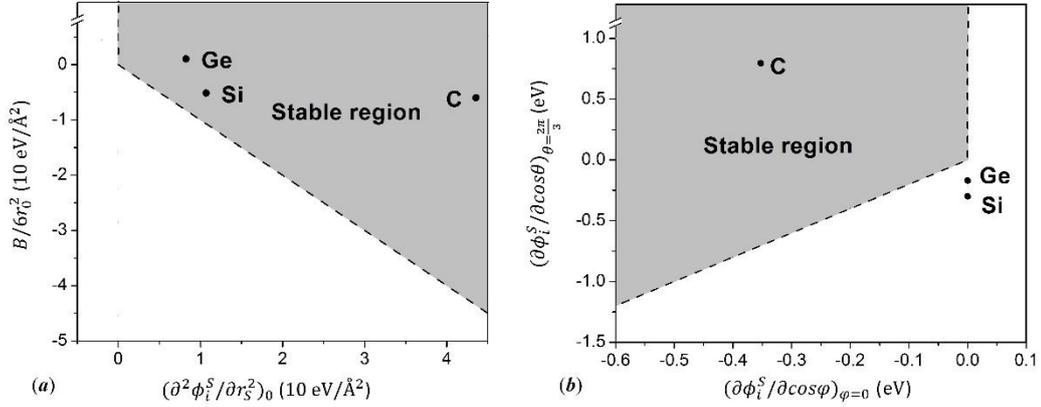

**Figure 2.** Phase diagrams of a planar honeycomb structure. (a) is derived from Eq. (25a), and (b) is derived from Eq. (25b). The relevant derivatives of interatomic potential for carbon (Brenner et al., 2002), silicon (Erhart and Albe, 2005), and germanium (Tersoff, 1989) are marked by black dots. The gray region corresponds to the stable region. The unstable regions are the white region and the edge of the gray

region, which are derived by positive non-positive definiteness and semi-definiteness of $[\mathbf{C}]$ or $[\mathbf{D}]$, respectively.

A square structure is a simple structure. The square structure and the related angles of interaction between atom $i$ and atom $j$ are shown in Fig.1(c).

The forms of $[\mathbf{C}]$ and $[\mathbf{D}]$ in a square structure are the same as those in a honeycomb, and using the symmetry in a square structure, we obtain:

$$C_{11} = C_{22} = \frac{1}{2}\left[\sum_{l=1}^{m}\sum_{l}\left(\frac{\partial^2 \phi_i^l}{\partial r_l^2}r_l - \frac{\partial \phi_i^l}{\partial r_l}\right)r_l\left(n_x^l\right)^4\right]_0,$$

$$C_{12} = \frac{1}{2}\left[\sum_{l=1}^{m}\sum_{l}\left(\frac{\partial^2 \phi_i^l}{\partial r_l^2}r_l - \frac{\partial \phi_i^l}{\partial r_l}\right)r_l\left(n_x^l\right)^2\left(n_y^l\right)^2\right]_0, \quad (26a)$$

$$C_{33} = 32\left(\frac{\partial^2 \phi_i^S}{\partial \cos\theta^2} - \frac{\partial^2 \phi_i^S}{\partial \cos\theta \partial \cos\theta'}\right)_{\theta,\theta'=\frac{\pi}{2}} + 2\left[\sum_{l=1}^{m}\sum_{l}\left(\frac{\partial^2 \phi_i^l}{\partial r_l^2}r_l - \frac{\partial \phi_i^l}{\partial r_l}\right)r_l\left(n_x^l\right)^2\left(n_y^l\right)^2\right]_0,$$

and

$$D_{11} = D_{22} = 2r_0^2\left[\left(\frac{\partial^2 \phi_i^S}{\partial \cos\theta}\right)_{\theta=\pi} - \left(\frac{\partial \phi_i^S}{\partial \cos\varphi}\right)_0\right] - \frac{1}{24}\left[\sum_{l=1}^{m}\sum_{l}\left(\frac{\partial \phi_i^l}{\partial r_l}\right)r_l^3\left(n_x^l\right)^4\right]_0,$$

$$D_{12} = 2r_0^2\left(\frac{\partial \phi_i^S}{\partial \cos\theta}\right)_{\theta=\frac{\pi}{2}} - \frac{1}{24}\left[\sum_{l=1}^{m}\sum_{l}\left(\frac{\partial \phi_i^l}{\partial r_l}\right)r_l^3\left(n_x^l\right)^2\left(n_y^l\right)^2\right]_0, \quad (26b)$$

$$D_{33} = -8r_0^2\left(\frac{\partial \phi_i^S}{\partial \cos\varphi}\right)_{\varphi=0} - \frac{1}{6}\left[\sum_{l=1}^{m}\sum_{l}\left(\frac{\partial \phi_i^l}{\partial r_l}\right)r_l^3\left(n_x^l\right)^2\left(n_y^l\right)^2\right]_0.$$

Similarly, when the potential of a square lattice only possesses short-range interaction, we have

$$C_{11} = C_{22} = r_0^2\left(\frac{\partial^2 \phi_i^S}{\partial r_S^2}\right)_0, \quad C_{12} = 0,$$

$$C_{33} = 32\left(\frac{\partial^2 \phi_i^S}{\partial \cos\theta^2} - \frac{\partial^2 \phi_i^S}{\partial \cos\theta \partial \cos\theta'}\right)_{\theta,\theta'=\frac{\pi}{2}}, \quad (27a)$$

$$D_{11} = D_{22} = 2r_0^2 \left[ \left( \frac{\partial^2 \phi_i^S}{\partial \cos\theta} \right)_{\theta=\pi} - \left( \frac{\partial \phi_i^S}{\partial \cos\varphi} \right)_{\varphi=0} \right],$$

$$D_{12} = 2r_0^2 \left( \frac{\partial \phi_i^S}{\partial \cos\theta} \right)_{\theta=\frac{\pi}{2}}, \quad D_{33} = -8r_0^2 \left( \frac{\partial \phi_i^S}{\partial \cos\varphi} \right)_{\varphi=0}, \tag{27b}$$

It could be defined as:

$$C' = \left( \frac{\partial^2 \phi_i^S}{\partial \cos\theta^2} - \frac{\partial^2 \phi_i^S}{\partial \cos\theta \partial \cos\theta'} \right)_{\theta,\theta'=\frac{\pi}{2}}. \tag{28}$$

Hence, the mechanical stability conditions of 2-D square structure materials are:

$$\left( \frac{\partial^2 \phi_i^S}{\partial r_S^2} \right)_0 > 0, \tag{29a}$$

$$C' > 0,$$

$$\left( \frac{\partial^2 \phi_i^S}{\partial \cos\theta} \right)_{\theta=\pi} + \left( \frac{\partial \phi_i^S}{\partial \cos\theta} \right)_{\theta=\frac{\pi}{2}} - \left( \frac{\partial \phi_i^S}{\partial \cos\varphi} \right)_{\varphi=0} > 0,$$

$$\left( \frac{\partial^2 \phi_i^S}{\partial \cos\theta} \right)_{\theta=\pi} - \left( \frac{\partial \phi_i^S}{\partial \cos\theta} \right)_{\theta=\frac{\pi}{2}} - \left( \frac{\partial \phi_i^S}{\partial \cos\varphi} \right)_{\varphi=0} > 0, \tag{29b}$$

$$\left( \frac{\partial \phi_i^S}{\partial \cos\varphi} \right)_{\varphi=0} < 0.$$

The phase diagrams derived from Eq. (29) are shown in Fig. 3.

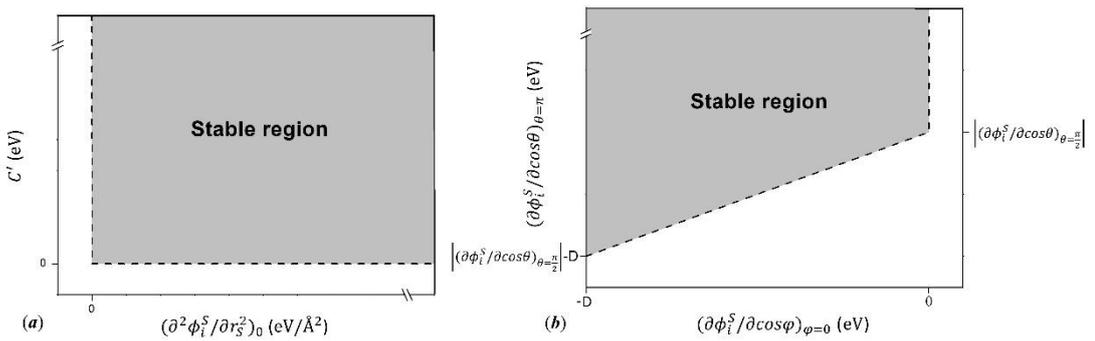

**Figure 3.** Phase diagrams of a planar square structure. (a) is derived from Eq. (29a), (b) is derived from Eq. (29b).

A triangular structure is also a simple structure. The triangular structure and the related angles of interaction between atom $i$ and atom $j$ are shown in Fig.1(d).

Similarly, $[C]$ and $[D]$ have the same forms of Eq. (19), and the matrix elements are:

$$C_{11} = C_{22} = \frac{3}{32}C_0 + \frac{1}{2}\left[\sum_{l=1}^{m}\sum_{l}\left(\frac{\partial^2 \phi_i^l}{\partial r_l^2}\right)r_l^2 (n_x^l)^4\right]_0,$$

$$C_{12} = -\frac{3}{32}C_0 + \frac{1}{2}\left[\sum_{l=1}^{m}\sum_{l}\left(\frac{\partial^2 \phi_i^l}{\partial r_l^2}\right)r_l^2 (n_x^l)^2 (n_y^l)^2\right]_0, \quad (30a)$$

$$C_{33} = \frac{3}{8}C_0 + 2\left[\sum_{l=1}^{m}\sum_{l}\left(\frac{\partial^2 \phi_i^l}{\partial r_l^2}\right)r_l^2 (n_x^l)^2 (n_y^l)^2\right]_0,$$

$$D_{11} = D_{22} = D_1 - \frac{1}{24}\left[\sum_{l=1}^{m}\sum_{l}\left(\frac{\partial \phi_i^l}{\partial r_l}\right)r_l^3 (n_x^l)^4\right]_0,$$

$$D_{12} = D_2 - \frac{1}{24}\left[\sum_{l=1}^{m}\sum_{l}\left(\frac{\partial \phi_i^l}{\partial r_l}\right)r_l^3 (n_x^l)^2 (n_y^l)^2\right]_0, \quad (30b)$$

$$D_{33} = D_3 - \frac{1}{6}\left[\sum_{l=1}^{m}\sum_{l}\left(\frac{\partial \phi_i^l}{\partial r_l}\right)r_l^3 (n_x^l)^2 (n_y^l)^2\right]_0,$$

where

$$C_0 = 3\left\{2\left[\left(\frac{\partial \phi}{\partial \cos\theta}\right)_{\theta=\frac{2\pi}{3}} - \left(\frac{\partial \phi}{\partial \cos\theta}\right)_{\theta=\frac{\pi}{3}}\right] + 3\left[\left(\frac{\partial^2 \phi}{\partial \cos\theta^2}\right)_{\theta=\frac{2\pi}{3}} + \left(\frac{\partial^2 \phi}{\partial \cos\theta^2}\right)_{\theta=\frac{\pi}{3}}\right]\right.$$
$$\left. -6\left(\frac{\partial^2 \phi}{\partial \cos\theta \partial \cos\theta'}\right)_{\theta=\frac{2\pi}{3},\theta'=\frac{\pi}{3}} -2r_0\left[\left(\frac{\partial^2 \phi}{\partial r \partial \cos\theta}\right)_{\theta=\frac{2\pi}{3}} - \left(\frac{\partial^2 \phi}{\partial r \partial \cos\theta}\right)_{\theta=\frac{\pi}{3}}\right]\right\}, \quad (31a)$$

$$D_1 = r_0^2\left[\frac{1}{4}D_1' + \frac{3}{4}D_2' - 21\left(\frac{\partial \phi_i^S}{\partial \cos\varphi}\right)_{\varphi=0}\right],$$

$$D_2 = r_0^2\left[\frac{1}{4}D_1' + \frac{1}{4}D_2' + \left(\frac{\partial \phi_i^S}{\partial \cos\varphi}\right)_{\varphi=0}\right], \quad (31b)$$

$$D_3 = r_0^2\left[D_2' - 44\left(\frac{\partial \phi_i^S}{\partial \cos\varphi}\right)_{\varphi=0}\right],$$

and

$$D_1' = 9\left[\left(\frac{\partial^2 \phi_i^S}{\partial \cos\theta}\right)_{\theta=\frac{\pi}{3}} + \left(\frac{\partial^2 \phi_i^S}{\partial \cos\theta_{ijk}}\right)_{\theta=\frac{2\pi}{3}}\right],$$

$$D_2' = 3\left[\left(\frac{\partial^2 \phi_i^S}{\partial \cos\theta}\right)_{\theta=\pi} - \left(\frac{\partial^2 \phi_i^S}{\partial \cos\theta}\right)_{\theta=\frac{\pi}{3}}\right].$$

(31c)

When the interatomic potential only possesses short-range interaction, we have

$$C_{11} = C_{22} = \frac{3}{32}r_0^2\left[3\left(\frac{\partial^2 \phi_i^S}{\partial r_S^2}\right)_0 + \frac{C_0}{r_0^2}\right],$$

$$C_{12} = \frac{3}{32}r_0^2\left[\left(\frac{\partial^2 \phi_i^S}{\partial r_S^2}\right)_0 - \frac{C_0}{r_0^2}\right], \quad C_{33} = \frac{3}{8}r_0^2\left[\left(\frac{\partial^2 \phi_i^S}{\partial r_S^2}\right)_0 + \frac{C_0}{r_0^2}\right],$$

(32a)

$$D_{11} = D_{22} = r_0^2\left[\frac{1}{4}D_1' + \frac{3}{4}D_2' - 21\left(\frac{\partial \phi_i^S}{\partial \cos\varphi}\right)_{\varphi=0}\right],$$

$$D_{12} = r_0^2\left[\frac{1}{4}D_1' + \frac{1}{4}D_2' + \left(\frac{\partial \phi_i^S}{\partial \cos\varphi}\right)_{\varphi=0}\right],$$

(32b)

$$D_{33} = r_0^2\left[D_2' - 44\left(\frac{\partial \phi_i^S}{\partial \cos\varphi}\right)_{\varphi=0}\right].$$

The mechanical stability conditions are:

$$\left(\frac{\partial^2 \phi_i^S}{\partial r_S^2}\right)_0 > 0,$$

$$\left(\frac{\partial^2 \phi_i^S}{\partial r_S^2}\right)_0 + \frac{C_0}{r_0^2} > 0,$$

(33a)

$$D_1' + 2D_2' - 40\left(\frac{\partial \phi_i^S}{\partial \cos\varphi}\right)_{\varphi=0} > 0,$$

$$D_2' - 44\left(\frac{\partial \phi_i^S}{\partial \cos\varphi}\right)_{\varphi=0} > 0.$$

(33b)

The phase diagrams derived from Eq. (33) are shown in Fig. 4.

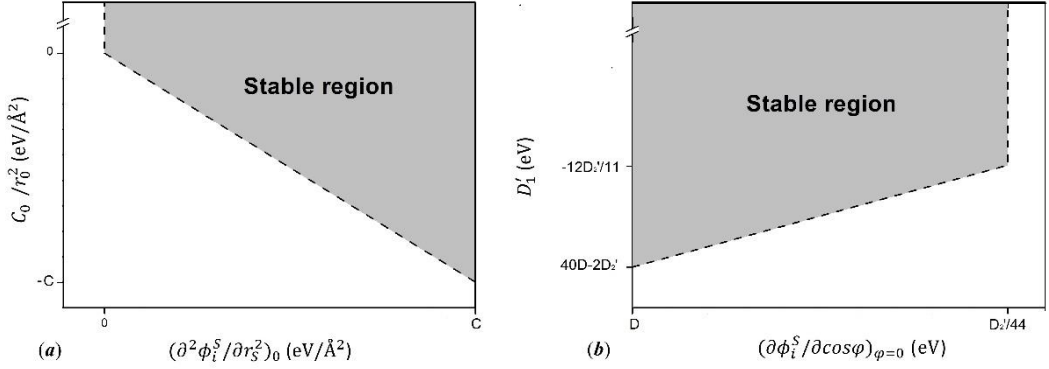

**Figure 4.** Phase diagrams of a triangular square structure. (a) is derived from Eq. (33a), (b) is derived from Eq. (33b).

## 4. Existence criterion for low-D materials described by typical atomistic potential

### 4.1.1. Results for pair potential

The pair potential, which only depends on the radial distance of two interacting atoms, possesses the simplest form and it could describe the interatomic interaction of some molecular crystals and atomic crystals well. In the case of pair potential, the bond angle and torsion angle terms in Eq. (2) vanish, which results in $\frac{\partial \phi_i^S}{\partial \cos\theta} \equiv 0$ and $\frac{\partial \phi_i^S}{\partial \cos\varphi} \equiv 0$. We define the equilibrium distance of pair potential as $r_p$, i.e., $\frac{\partial \phi_i^l}{\partial r_l} = 0$ for $r_l = r_p$, and $\frac{\partial \phi_i^l}{\partial r_l} > 0$ for $r_l > r_p$. As the first neighbor interaction $\phi_i^S$ makes the dominant contribution to total potential, the equilibrium bond length $r_0$ must be near to $r_p$. As a result, the $l$th-nearest neighbor distance $r_l > r_p$ ($l \geq 2$), and we have $\left(\frac{\partial \phi_i^l}{\partial r_l}\right)_0 > 0$. Here, we qualitatively prove that pair potential is inappropriate to describe the physical properties of straight/planar low-D equal-bond-length materials.

For 1-D monatomic chains, Eq. (17b) can be converted to

$$\sum_{l=1}^{m}\sum_{l}\left(\frac{\partial \phi_i^l}{\partial r_l}\right)_0 (r_l^3)_0 < 0 \tag{34}$$

in this case.

Note that Eq. (11a) can be rewritten as

$$\sum_{l=1}^{m}\sum_{l}\left(\frac{\partial \phi_i^l}{\partial r_l}\right)_0 (r_l)_0 = 0. \tag{35}$$

According to Eq. (35), as $\left(\frac{\partial \phi_i^l}{\partial r_l}\right)_0 > 0$ when $l \geq 2$, one can immediately obtain:

$$\sum_{l=1}^{m}\sum_{l}\left(\frac{\partial \phi_i^l}{\partial r_l}\right)_0 (r_l^3)_0 > 0. \tag{36}$$

This result contradicts the inequation in Eq. (34). It means the pair potential cannot be used to investigate 1-D monatomic chains.

For the three permissible planar 2-D, equal-bond-length lattice structures, the inequations derived from the positive definiteness of $[\mathbf{D}]$ have the same forms

$$\begin{aligned}
\sum_{l=1}^{m}\sum_{l}\left(\frac{\partial \phi_i^l}{\partial r_l}\right)_0 (r_l^3)_0 \left[(n_x^l)^4 - (n_x^l)^2 (n_y^l)^2\right] &< 0, \\
\sum_{l=1}^{m}\sum_{l}\left(\frac{\partial \phi_i^l}{\partial r_l}\right)_0 (r_l^3)_0 \left[(n_x^l)^4 + (n_x^l)^2 (n_y^l)^2\right] &< 0, \\
\sum_{l=1}^{m}\sum_{l}\left(\frac{\partial \phi_i^l}{\partial r_l}\right)_0 (r_l^3)_0 (n_x^l)^2 (n_y^l)^2 &< 0.
\end{aligned} \tag{37}$$

Obviously, these inequations are not independent: the second one can be derived by a linear combination of the first and third ones. Owing to the symmetry of the three lattice structures, Eq. (12a) can be rewritten as

$$\sum_{l=1}^{m}\sum_{l}\left(\frac{\partial \phi_i^l}{\partial r_l}\right)_0 (r_l)_0 \left[\left(n_x^l\right)^4 + \left(n_x^l\right)^2\left(n_y^l\right)^2\right] = 0. \tag{38}$$

Similarly, $\left(\frac{\partial \phi_i^l}{\partial r_l}\right)_0 > 0$ when $l \geq 2$, we have

$$\sum_{l=1}^{m}\sum_{l}\left(\frac{\partial \phi_i^l}{\partial r_l}\right)_0 \left(r_l^3\right)_0 \left[\left(n_x^l\right)^4 + \left(n_x^l\right)^2\left(n_y^l\right)^2\right] > 0 \tag{39}$$

This result contradicts the inequation in Eq. (38), namely, $[\mathbf{D}]$ is non-positive definite for any pair potential planar 2-D, equal-bond-length materials.

As a result, by only considering pair potential, the straight/planar, low-D, equal-bond-length materials are unstable. In this sense, these materials can only be understood by assuming multi-body interatomic potentials.

### 4.1.2. Results for the REBO potential

The REBO potential is one of the effective analytical multi-body potentials. Despite its lack of long-range interactions, the REBO potential predicts rather accurate value for lattice constant, force constants, and elastic constants of diamond and graphite. The general formulation of the REBO potential is (Harrison et al., 2015)

$$\phi_i^{REBO} = \sum_{j}^{Neighbour} \phi_{ij},$$
$$\phi_{ij} = f_c(r_{ij})[\phi^R(r_{ij}) - B_{ij}(\{cos\theta\},\{cos\varphi\})\phi^A(r_{ij})], \tag{40}$$

where $\phi_{ij}$ is the interaction energy between the neighbor atoms $i$ and $j$, and $r_{ij}$ is the distance between the two atoms. $f_c(r_{ij})$ is defined by a switching function. $\phi^R$ and $\phi^A$ are pair-additive repulsive and attractive interactions, respectively. $B_{ij}$ is the total bond order between atoms $i$ and $j$, and it can be further written as

$$B_{ij} = \frac{b_{ij} + b_{ji}}{2} + \Pi_{ij} + T_{ij}, \tag{41}$$

where $b_{ij}$ and $b_{ji}$ are covalent bonding terms, and for elemental materials, it gives

$$b_{ij} = \left[1 + \sum_{k \neq i,j} G(cos\theta_{ijk})\right]^{-\frac{1}{2}}, \quad (42)$$

where $G(cos\theta_{ijk})$ represents the contribution of bond angles to the covalent bonding term, and it can be expressed as a six-order spline function. $\Pi_{ij}$ is a conjugation term that depends on local conjugation, and $T_{ij}$ is a torsion term

$$T_{ij} = T_0 \left[\sum_{k \neq i,j} \sum_{l \neq i,j} (1 - \cos^2 \varphi_{ijkl})\right], \quad (43)$$

where $T_0$ depends on the coordinate number.

By assuming REBO potential, the mechanical stability condition of 1-D monatomic chains in Eq. (17) is:

$$\phi^{R''}(r_0) - B_{ij}\phi^{A''}(r_0) > 0, \quad (44a)$$

$$b_{ij}^3\left(-1, -\frac{1}{2}\right) G'(-1) > 0. \quad (44b)$$

Since the interatomic potential increases rapidly when atoms deviate away from the equilibrium position, $\phi^{R''}(r_0) - B_{ij}\phi^{A''}(r_0)$ is a relatively large positive value. The inequations in Eq. (44a) could be true.

As $G(cos\theta_{ijk})$ is not an analytical function in REBO potential, the properties of $G(cos\theta_{ijk})$ would be illustrated qualitatively. It is reasonable that a smaller angle has a much greater influence on bond order than a larger one. Hence $G(cos\theta_{ijk})$ should be a monotonic increasing function, which means $G'(cos\theta_{ijk}) > 0$. Since $b_{ij}(\{\cos\theta_{ijk}\}) > 0$, inequation in Eq. (44b) could be also true. Therefore, the 1-D monatomic chains could be stable.

In the case of REBO potential, the mechanical stability conditions for the planar honeycomb, square, and triangular structures are shown in Eqs. (25), (29), and (33), respectively.

For a honeycomb structure, Eq. (25) yields

$$\phi^{R''}(r_0) - B_{ij}\phi^{A''}(r_0) > 0$$
$$\phi^{R''}(r_0) - B_{ij}\phi^{A''}(r_0) + \frac{B}{6r_0} > 0 \quad , \tag{45a}$$

$$b_{ij}^3\left(-\frac{1}{2}, -\frac{1}{2}\right)G'\left(-\frac{1}{2}\right) > 16T_0$$
$$T_0 < 0 \tag{45b}$$

As demonstrated above, $\phi^{R''}(r_0) - B_{ij}\phi^{A''}(r_0)$ is a relatively large positive value. The inequations in Eq. (45a) could be true.

$T_0 < 0$ implies the energy of a planar configuration is lower than that of twisty configuration, and it is the basic requirement for a planar lattice. Since $b_{ij}(\{\cos\theta_{ijk}\}) > 0$ and $G'(\cos\theta_{ijk}) > 0$, the conditions in Eq. (45b) could be satisfied simultaneously. Therefore, the planar honeycomb structure could be stable.

For a square structure, the inequations in Eq. (29) can be converted to

$$\phi^{R''}(r_0) - B_{ij}\phi^{A''}(r_0) > 0$$
$$G''(0) > 0 \quad , \tag{46a}$$

$$b_{ij}^3(0,0,-1)\left[G'(-1) + G'(0)\right] > 8T_0$$
$$b_{ij}^3(0,0,-1)\left[G'(-1) - G'(0)\right] > 8T_0 . \tag{46b}$$
$$T_0 < 0$$

According to the parameters for the angular contribution to the bond order for carbon (Brenner et al., 2002) and silicon (Cahangirov et al., 2009) bonding, we assume that, for many covalent systems, $G'(\cos\theta_{ijk})$ would also be a monotonic increasing function and it implies $G'(-1) - G'(0) < 0$. However, $|G'(-1) - G'(0)|$ is an order of magnitude larger than $|T_0|$, due to the relatively low torsional barrier (Los et al., 2005).

Thus, the inequations in Eq. (46b) could contradict each other. The planar square structure could not stably exist.

For a triangular structure, the conditions in Eq. (33) can be converted to

$$\phi^{R\prime\prime}(r_0) - B_{ij}\phi^{A\prime\prime}(r_0) > 0 \\ \phi^{R\prime\prime}(r_0) - B_{ij}\phi^{A\prime\prime}(r_0) + C_0 > 0 \quad (47a)$$

$$b_{ij}^3\left(\frac{1}{2},\frac{1}{2},-\frac{1}{2},-\frac{1}{2},-1\right)\left[G'\left(\frac{1}{2}\right)+3G'\left(-\frac{1}{2}\right)+2G'(-1)\right] > \frac{320}{3}T_0 \\ b_{ij}^3\left(\frac{1}{2},\frac{1}{2},-\frac{1}{2},-\frac{1}{2},-1\right)\left[G'(-1)-G'\left(\frac{1}{2}\right)\right] > \frac{352}{3}T_0 \quad (47b)$$

It seems that the inequations in Eq. (47) can be satisfied simultaneously. However, the coordinate number in planar configuration is no more than 4 because of the directivity and saturability of covalent bonding (Pauling, 1960). Since the coordinate number of a planar triangular structure is 6, the 2-D structure cannot exist.

Besides, when the coordinate number increases, the bonds are weakened, which results in bond length increases. The bond length of a planar triangular structure seems to be relatively long. Take carbon bonding as an example. The bond length of a planar carbon triangular lattice is $r_0 = 2.10 \text{Å}$ for the REBO potential. However, the maximum cutoff parameter of the switching function $f_c(r_{ij})$ is $2.0 \text{Å}$, the carbon atoms in a triangular lattice possess no cohesion, and thus it is not stable.

Therefore, according to the REBO potential, a 1-D monatomic chain and the planar honeycomb structure are the most stable configurations.

## 5. Tentative exploration for 2-D honeycomb-like buckled structure

Most 2-D materials discovered at present are not in planar forms (Balendhran et al.,

2015; O'hare et al., 2012; Sahin et al., 2009). As mentioned above, the stable low-D structures for silicon and germanium are the honeycomb-like buckled structures, and they are respectively called silicene and germanene. In this structure, the neighboring atoms are located in different parallel planes with the distance between these two planes being $2\delta$ (Fig. 5).

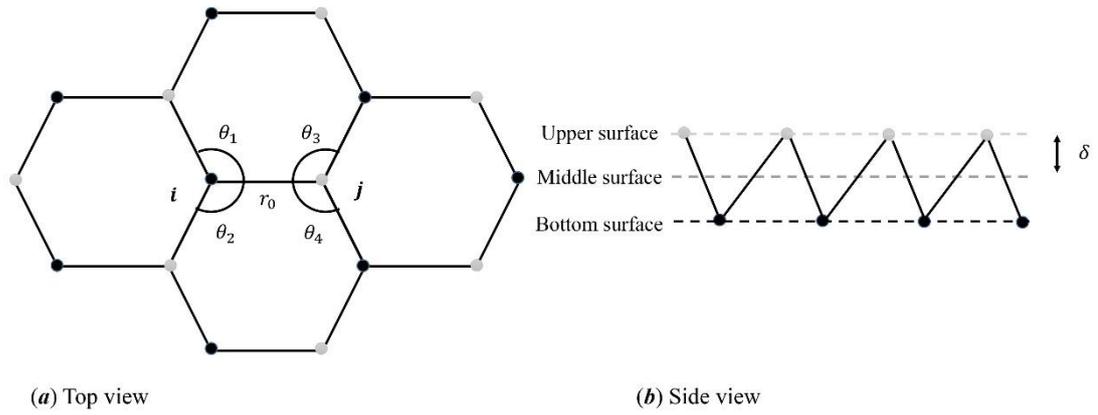

(a) Top view  (b) Side view

**Figure 5.** The honeycomb-like buckled structure

Similarly, we explored the existence criterion of a honeycomb-like buckled structure. Assuming that the interatomic potential only possesses short-range interaction and the torsion angle term vanishes, the total interatomic potential is:

$$\Phi = \frac{1}{2} N \sum_S \phi_i^S \left( r_S, \{\cos\theta\} \right), \tag{48}$$

where the initial value of $r_{ij}$ is $\sqrt{r_0^2 + 4\delta^2}$.

An extremal $\Phi$ requires $\left(\frac{\partial \Phi}{\partial r_S}\right)_0 = \left(\frac{\partial \Phi}{\partial \cos\theta}\right)_0 = 0$, because there is no geometric constraint for $r_{ij}$ and $\theta_{ijk}$. In this case, the initial radius of curvature is comparable to the atomic spacing when using the description of a single smooth curved surface. Therefore, Eq. (5) is no longer applicable for the buckled structure. From the viewpoint of differential geometry, the honeycomb-like buckled structure is consists of three surfaces: the upper surface, the middle surface and the bottom surface (Fig. 5(b)). The upper surface contains the all the atoms of one of the sub-lattices, the bottom one contains the rest, and the middle one is considered as a reference surface. Then the deformed $r_{ij}$ and $cos\theta_{ijk}$ can be derived. Their derivation process is described in the Appendix.

According to the expressions of deformed $r_{ij}$ and $cos\theta_{ijk}$, it can be derived that strain and curvature are also decoupled in energy. Thus, matrix $[M]$ is also block diagonal, and then the criterion is derived by the positive definiteness of $[C]$ and $[D]$. Because the analytical form of $[C]$ and $[D]$ of a honeycomb-like buckled structure is complicated, the optimizing structural parameters and numerical value of elements of $[C]$ and $[D]$ for silicene and germanene, which derived by the Tersoff-type potential for silicon (Erhart and Albe, 2005) and germanium (Tersoff, 1989), are shown in Table 1.

|  | $r_0$ (Å) | $\delta$ (Å) | $C_{11}$ (eV) | $C_{12}$ (eV) | $C_{33}$ (eV) | $D_{11}$ (eV·Å²) | $D_{12}$ (eV·Å²) | $D_{33}$ (eV·Å²) |
|---|---|---|---|---|---|---|---|---|
| Silicene | 2.18 | 0.38 | 27.58 | 8.46 | 38.23 | 5.20 | 1.73 | 6.93 |
| Germanene | 2.38 | 0.25 | 43.53 | -8.75 | 104.56 | 15.93 | 5.31 | 21.25 |

Table 1. The structural parameters and elements of $[C]$ and $[D]$ for silicene and germanene, where $C_{13}=C_{23}=0$ and $D_{13}=D_{23}=0$.

Clearly, both $[C]$ and $[D]$ for silicene and germanene are positive definite. Therefore, according to our method, the honeycomb-like buckled structures of silicon and germanium are stable rather than a planar configuration. This result is consistent with recent theories and experiments of silicene and germanene (Cahangirov et al., 2009;

Davila et al., 2014; Feng et al., 2012; O'hare et al., 2012).

## 6. Summary

To summarize, an atomistic-based existence criterion for low-D materials is established as follows: 1.Determining the permissible structures by deriving the extreme value condition of the potential energy $\Phi$ of the system. 2. Determining the mechanical stability of these structures by analyzing the positive definiteness of the block diagonal matrix $[\mathbf{M}]$, whose elements are the quadratic terms of the potential energy. For straight/planar, low-D, equal-bond-length elemental materials, $[\mathbf{M}]$ consists of two square matrixes $[\mathbf{C}]$ and $[\mathbf{D}]$, and $[\mathbf{D}]$ shows the leading feature of the stability of low-D materials. The criterion is expressed in the form of inequalities for several parameters of interatomic potential, and it varies depending on the lattice structure. Only a 1-D monatomic chain, and honeycomb, square, and triangular structures are permissible for straight/planar, low-D, equal-bond-length elemental materials. It is proved that pair potential cannot apply to these structures. Besides, by assuming the REBO potential, a 1-D monatomic chain and honeycomb structure are likely to be most stable, whereas a square structure and triangular structure could be unstable. We found that carbon can form a stable planar 2-D honeycomb structure, while silicon and germanium cannot. Yet, silicon and germanium can remain stable in a honeycomb-like buckled structure. Note that the criterion provides the first step of stable analysis. It can provide a basic guideline for scientists to choose and fabricate stable low-D materials, and it should be improved upon by considering the thermal vibration and external field effects, which is to be discussed in our future work.

**Acknowledgments**: This work was supported by the National Natural Science Foundation through the grants (11472313, 11232015, 11572355, 11502308).

**Author contributions**: B. Wang conceived the idea. J. Chen finished the analytical deduction. J. Chen, Y. Hu and B. Wang discussed the results for revision. J. Chen, Y. Hu and B. Wang co-wrote the manuscript. All authors reviewed the manuscript.

**Competing financial interests:** The authors declare no competing financial interests.

**Appendix. The deformed** $r_{ij}$ **and** $cos\theta_{ijk}$ **in a honeycomb-like buckled structure**

The deformed honeycomb-like buckled structure is shown in Fig. A.1. The strain and curvature of the buckled structure is defined as that of the middle surface. The projection points of points $i$ and $j$ onto the middle surface are $i'$ and $j'$.

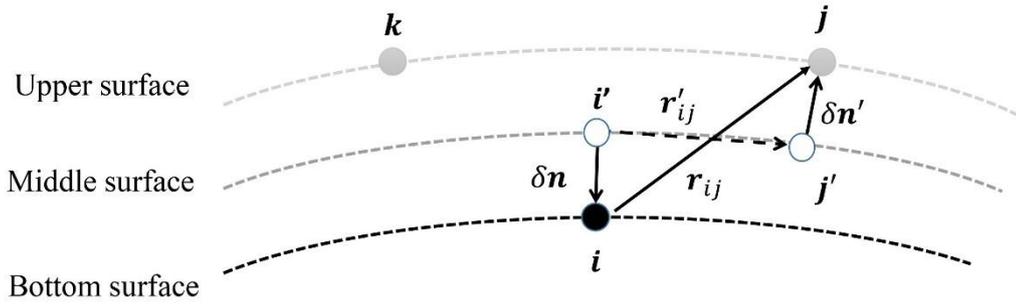

**Fig. A.1** The side view of a deformed honeycomb-like buckled structure

The vector between the two atoms $i$ and $j$ is:

$$\mathbf{r}_{ij} = \mathbf{r}_{ij}' + \delta \mathbf{N}' - \delta \mathbf{N}, \tag{A.1}$$

where $\mathbf{N}$ and $\mathbf{N}'$ are the unit normal vector at $i'$ and $j'$, respectively. In addition, $\mathbf{r}_{ij}'$ is the vector between $i'$ and $j'$.

For infinitesimal deformation, the form of $\mathbf{r}_{ij}'$ is shown in Eq. (5), where $n_\alpha$ and $n_\beta$ are the direction vector between $i'$ and $j'$ before deformation.

$\mathbf{N}'$ is obtained by Taylor series expansion:

$$\mathbf{N}' = \mathbf{N} + \frac{\partial \mathbf{N}}{\partial \alpha} r_0 (n_\alpha + x_\alpha) + \frac{1}{2} \frac{\partial^2 \mathbf{N}}{\partial \alpha \partial \beta} r_0 (n_\alpha + x_\alpha)(n_\beta + x_\beta). \tag{A.2}$$

The second- and third-order derivatives in Eq. (A.2) are:

$$\begin{aligned}\frac{\partial \mathbf{N}}{\partial \alpha} &= -\kappa_{\alpha\beta} T^{\beta\gamma} \mathbf{T}_\gamma \\ \frac{\partial \mathbf{N}}{\partial \alpha} &= -\kappa_{\alpha\gamma} \kappa_{\lambda\beta} T^{\gamma\lambda} \mathbf{N}\end{aligned}. \tag{A.3}$$

$\mathbf{N}'$ then becomes:

$$\begin{aligned}\mathbf{N}' = &\left[1 - \frac{1}{2} \kappa_{\alpha\gamma} \kappa_{\lambda\beta} T^{\gamma\lambda} r_0^2 (n_\alpha + x_\alpha)(n_\beta + x_\beta)\right] \mathbf{N} \\ &- \kappa_{\alpha\beta} T^{\beta\gamma} r_0 (n_\alpha + x_\alpha) \mathbf{T}_\gamma\end{aligned}. \tag{A.4}$$

Hence, $\mathbf{r}_{ij}$ is expressed by covariant base vectors $\mathbf{T}$ and the unit normal vector $\mathbf{N}$ at $i'$. The distance between atoms $i$ and $j$ is $r_{ij} = |\mathbf{r}_{ij}|$, and $cos\theta_{ijk} = \frac{\mathbf{r}_{ij} \mathbf{r}_{ik}}{|\mathbf{r}_{ij}||\mathbf{r}_{ik}|}$.